\documentclass[prb,aps,twocolumn,superscriptaddress]{revtex4}
\usepackage{graphicx}
\usepackage{amssymb}
\usepackage{bm}

\begin{document}

\author{Toshifumi Itakura}
 \affiliation{Nagoya Univ. Department of Phys., Nagoya 465-0054, Japan}
 \email{itakurat@s6.dion.ne.jp}
\author{Fumitada Itakura}
 \affiliation{Meijo Univ. Department of Engineer., Nagoya 465-0054, Japan}

\title{Decoherence of one-dimensional electron system}

\begin{abstract}
Recently, there have been many attempts to implement quantum computation experimentally.
For this purpose, quantum coherence should be maintained during gate operations.
Therefore, the control of decoherence is a very important problem.
In present study, we examine decoherence of one-dimensional electron system, which coupled with acoustic phonon.
The time-dependent Hatree-Fock approximation is applied for electron system, the Schrodinger equation is applied for phonon.
The metal-insulator transition found for coherence time.
The results are the electron interaction reduces the coherence time.
We also examined the Bang-Bang control, the results shows recovery of coherence.

\end{abstract}


\maketitle

\section{Introduction}

Among the various proposals for quantum computation, quantum bits (qubits) in solid state materials, such as superconducting Josephson junctions 
\cite{Nakamura} and quantum dots, \cite{Hayashi,Tanamoto,Loss} have the advantage of scalability. Proposals to implement a quantum computer using superconducting nanocircuits are proving to be very promising, and several experiments have already highlighted the quantum properties of these devices.
Such a coherent two-level system constitutes a qubit and the quantum computation can be carried out as a unitary operation functioning on the multiple-qubit system. 
Essentially, this quantum coherence must be maintained during computation.
However, it is difficult to avoid dephasing due to the system's interaction with its external environment. 
The dephasing is characterized by a dephasing time of $T_2$.
Various environments can cause dephasing.
Background charge fluctuations (BCFs) have been observed in diverse kinds of systems.
In nanoscale systems, BCFs are electrostatic potential fluctuations arising due to the dynamics of an electron, or a hole, on a charge trap. In particular, the charges in charge traps fluctuate (TLS) with the Lorentzian spectrum form, this is called random telegraph noise in the time domain.
\cite{Itakura_Tokura}
The random distribution of the positions of such dynamical charge traps and their time constants leads to BCF or 1/f noise.
In solid sate charge qubits, these BCFs result in a dynamical electric disturbance and, hence, dephasing.
It should be noted that this dephasing process does not mean the qubit being entangled with the environment, but rather that the stochastic evolution of the external classical filed is suppressing the density matrix elements of the qubit after averaging out over statistically distributed samples.
We had shown that BCFs are important channel of dephasing for charge qubit system.
In the present study, we investigate the effect of phonon on the two-qubit-gate operation.
To construct a controllable quantum computer, one requires the suppression of dephasing and an accurate universal quantum gate which consists of one qubit and two-qubit operations. Therefore, to address these manipulations, we examine the dephasing of a coupled qubit system, which is an experimentally current topic, and urgent to analyze what cause of dephasing is important in this system.
Reentry, the qubit system of coupled dots has been examined.
\cite{Itakura_Tokura_2}
However, the detailed study of 2-qubit system and many qubit systems does not achieved.
In this study we examine the 2-qubit system which coupled with acoustic phonon system.
We treat boundary condition as periodic.
The examined system is equivalent with the two site unit cell 
Hubbard model coupled with dimeried acoustic phonon system.
The electron phonon system has been studied extensibly.
We specially examine about the interaction between electron systems.
The spin decoherence in which the interaction is included has been studied theoretically.
\cite{Sousa_1,Itakura_1}
Decoherence for charge qubit is studied of the authors when environment is BCF.
\cite{Itakura_Tokura}
Present study treats the one dimensional system, therefore the importance exist in mesoscopic physics also.

\subsection{1D Hamiltonian}

The Hamiltonian is defined as
\begin{equation}
 H = \sum_{i,\sigma} h_{i,\sigma}^{F} ( c_{i,\sigma}^{\dagger} c_{i+1,\sigma}
+ c_{i+1,\sigma}^{\dagger} c_{i,\sigma} )+
 + \sum_{i,\sigma} h_{i,\sigma}^H n_{i,\sigma}.
\end{equation}
with the creation and destruction operator, the Hatree-Fock approximation
is given as below,
\begin{eqnarray}
h_{i,\sigma}^{F} &=& - t_0 + \alpha ( u[i+1] -u[i]), \nonumber \\
h_{i,\sigma}^{H} &=& \frac{U}{2} <n_{i,\sigma}> + \mu[i,\sigma]. 
\end{eqnarray}
In above model, we choose the periodic boundary condition with two unit cell.
$t_0$ is hopping energy of phonon, $U$ is electron one site repulsion energy, $\alpha$ the interaction strength electron and acoustic phonon.
The time evolution electron system is given by,
\begin{eqnarray}
\frac{d g[i][j]^{\sigma \sigma'}}{d t}
 &=&\frac{i}{\hbar}[ h_{i,\sigma}^H g[i][j]^{\sigma \sigma'}
      + h_{i,\sigma}^F g[i+1][j]^{\sigma \sigma'} \nonumber \\
      &+& h_{i-1,\sigma}^F g[i-1][j]^{\sigma \sigma'}
      \nonumber \\
      &-& h_{j,\sigma}^H g[i][j]^{\sigma,\sigma'}
       - h_{j-1,\sigma}^F g[i][j-1]^{\sigma,\sigma'}
      \nonumber \\
      &-& h_{j,\sigma}^F g[i][j+1]^{\sigma,\sigma'}].
\end{eqnarray}
The time evolution of phonon system is given as below,
\begin{equation}
\frac{d v[i]}{d t} = -k u[i] -a(f(g[i][j])), \frac{d u[i]}{d t} = v[i]
\end{equation}
There four phonon mode.
Here $i=0=2$ and $i=3=1$, this is periodic boundary condition.
This boundary condition is important to include the phonon effect.
$f(g[i][j])$ the back reaction from electron system.

The initial condition for electron system given as below,
\begin{eqnarray}
g[0][1]^{\sigma,\sigma'}&=& -1/n ,g[1][0]^{\sigma,\sigma'}=-1/n
\end{eqnarray}
We choose $1/n$-filling system and assume the off-diagonal coherence,
\begin{eqnarray}
g[i][i]^{\sigma,\sigma'} &=& 1/n. 
\end{eqnarray}
The initial condition for phonon system is given as,
\begin{equation}
u[i]=(-1)^(i/2) 0.1,v[i]=0.
\end{equation}
Above initial condition of electron system is from the ground state to four excited state.

\subsection{numerical results}

We solve above time-dependent Hatree-Fock equation numerically.
First we examine the half-filled electron system. ($n=2$)
The auto correlation function of 2$k_F$SDW is Lorentz form.
Thus we consider the following two models.
The auto correlation function of one impurity capacity ($C_0$), n-the capacity ($C$) and resistant system is given by.
\begin{equation}
 \frac{1}{\tau}
 + \frac{1}{\tau} 
  \frac{\tau}{\tau^2+ (\omega - \Delta)^2}
\end{equation}
Here, $\tau=(C_0/(\frac{C}{n} + c_0))^2)$ and $\omega$ is resonance frequency.
The sum about frequency results in the constant and weighted time constant, as integrating about the Lorentz function is $\pi$.
The $\tau$ indicates the back reaction of electron system from interaction.
This quantity is not only dielectric property but also decoherence rate.
If auto correlation function is Lorentz form, we can examine easily decoherence rate and the phase transitions also.
Another estimation of auto correlation function is the coupled dot system which interacts with leads by tunneling effect.
By using linear response theory the expression of conductance is obtained as follows,
\begin{eqnarray}
\sigma ( \omega ) & \simeq & \frac{1}{\Gamma} \{ J (\omega), J(-\omega) \}
=\frac{1}{\Gamma}
\{ \frac{d n( \omega)}{d t}, \frac{d n (-\omega)}{d t} \} \nonumber \\
&=& \frac{1}{\Gamma} \frac{ i \omega}{ i (\omega - \Delta) + \Gamma} 
\frac{ - i \omega}{ - i (\omega - \Delta) + \Gamma} \nonumber \\ 
&=& \frac{\omega^2}{\Gamma((\omega - \Delta)^2 + \Gamma^2)} \nonumber \\
&=&
\frac{1}{\Gamma} + \frac{1}{\Gamma} 
\frac{\Gamma^2}{(\omega - \Delta)^2 +\Gamma^2}
\end{eqnarray}
where $\sigma ( \omega )$ is conductance, $\omega$ is frequency, 
$\Delta$ is Rabi oscillation frequency and $\Gamma$ is life time of oscillation.
We examine the order parameter's time constant by the use of above method.
The time constant of 2$k_F$ SDW shows transition at $U$=0.
The fock term that does not vanish represents the DC conductivity.
The fock term of charge decreases when transition.
That of spin ferromagnetic also decreases when transition.
That of spin singlet and triplet increases at transition.
Next we examine the collision term.
The collision term strongly depends on the sample time and simples mesh.
For $t=4096$ and $\delta t = 0.01$, the collision term for charge decreases and oscillation as a function of $U$.
The collision term for spin is increases with $U$ and 0 for $U=0$.
These results because the umklapp process and electron-phonon interaction.
The whole system is conserved system which interacts with electron conserved system and phonon conserved system.
The interaction induced decoherence occurs.

The auto correlation function shows single peak for small $U$ and several peaks for large $U$.
This behavior shows expanding limit cycle to quantum chaos.
The auto-correlation function of order parameter shows quantum chaos.
In Fourier space the $T_2$ torus appears for metallic side, the bifurcation to $T_3$ torus appears for insulator side.
The Ruelle-Takens type chaos appears.

In this study, we include the effect of spin off-diagonal element.
This effect is important due to the back reaction.

Next, we examine the spectrum of phonon degree of freedom.
The parameter dependence of the electron-phonon interaction shows the monotonic decreasing function. 
Decoherence disappears for zero electron-phonon interaction.
This is because the each subsystem is closed conserved system.
And due to the low dimensional fluctuation the structural phase transition does not appear.

The 3D plot of real part and imaginary part of order parameter and phonon displacement show complex periodic structure.
For $t_0=1,U=0.01$, the phase is metallic and periodic.
For $t_0=1,U=1.0$, the phase is insulator and chaotic.
To examine the recovery of coherence, we study the Bang-Bang control.
The effect of Bang-Bang control the metallic side becomes more periodic and insulter side becomes squeezed.
Thus the recovery of coherence by Bang-Bang control is observed.

The Bang-Bang control does the binary change of total wave function phase.
The recovery of coherence appears, in spite of that the time trajectory show oscillation by external force.
The Fourier transforms of auto correlation function shows periodic many bifurcation.
With changing the time constant of binary Bang-Bang control, one can do control of decoherence.
The Fourier transform of auto-correlation function consists of $T_2$ torus and bifurcated high frequency mode.
We propose this induced periodic structure for the photo-induced ferroelectric state.

\end{document}